\documentclass[aps,prl,reprint]{revtex4-1}
\usepackage[utf8]{inputenc}
\usepackage{graphicx}
\usepackage{textcomp}
\usepackage{amsmath}
\usepackage{hyperref}

\begin{document}

\title{A Quantum Network Node with Crossed Optical Fibre Cavities}

\author{Manuel Brekenfeld}
\email[To whom correspondence should be addressed. Email: ]{manuel.brekenfeld@mpq.mpg.de}
\author{Dominik Niemietz}
\author{Joseph Dale Christesen}
\altaffiliation[Present address: ]{National Institute of Standards and Technology, Boulder, USA}
\author{Gerhard Rempe}
\affiliation{Max-Planck-Institut für Quantenoptik, Hans-Kopfermann-Strasse 1, 85748 Garching, Germany}

\maketitle

\textbf{Quantum networks provide unique possibilities for resolving open questions on entanglement \cite{Acin2007} and promise innovative applications ranging from secure communication to scalable computation \cite{Wehner2018}. While two quantum nodes coupled by a single channel are adequate for basic quantum communication tasks between two parties \cite{Ritter2012}, fully functional large-scale quantum networks require a web-like architecture with multiply connected nodes \cite{Kimble2008}. Efficient interfaces between network nodes and channels can be implemented with optical cavities \cite{Reiserer2015}. Using two optical fibre cavities coupled to one atom, we here realise a quantum network node that connects to two quantum channels. It functions as a passive, heralded and high-fidelity quantum memory that requires neither amplitude- and phase-critical control fields \cite{Gorshkov2007,Specht2011,Koerber2018} nor error-prone feedback loops \cite{Kalb2015}. Our node is robust, fits naturally into larger fibre-based networks, can be scaled to more cavities, and thus provides clear perspectives for a quantum internet including qubit controlled quantum switches \cite{Reiserer2014,Tiecke2014}, routers \cite{Shomroni2014,Scheucher2016}, and repeaters \cite{Briegel1998,Uphoff2016}.} 

An essential requirement of a quantum network node is its capability to serve as a quantum memory that receives, stores and releases an unknown and potentially entangled quantum state better than any classical memory. Quantum memories have been implemented in various physical systems ranging from atoms to solids, and from ensembles to single emitters \cite{Bussieres2013}. Despite progress, a large challenge concerns the always present photon loss and the always finite efficiency. Both limitations require the addition of a herald that signals successful operation of the quantum memory node. Although several protocols with heralds have been proposed and investigated \cite{Lin2009, Koshino2010, Tanji2009, Kurz2014, Yang2016, Delteil2017, Bechler2018}, an advantage over a classical memory could be shown only in a few cases \cite{Chen2008, Kalb2015}. One class of realisations is based on quantum teleportation of a photonic qubit into a quantum memory, where the photonic Bell state measurement of the teleportation protocol provides the herald for successful storage \cite{Chen2008}. This scheme, however, faces a fundamental efficiency limit of 50\,\% and requires two indistinguishable photons, which is a significant challenge for any practical implementation. A new class arises when the quantum memory is coupled to an optical cavity. This enabled the heralded storage of a photonic qubit in a single atom by reflection of a photon from the cavity \cite{Kalb2015}. However, that memory was limited by two fundamental aspects of the protocol. First, the scheme requires active feedback onto the spin of the atom conditioned on the detection of the reflected photon, leading to extra experimental steps in the protocol that are prone to errors. Second, the detection of the reflected photon from the cavity only heralds the presence of a photon and not successful storage. Successful storage requires a well prepared atom and perfect coupling between the incoming photon and the cavity mode, neither of which are guaranteed. Accordingly, a storage fidelity comparable to that of other storage schemes, such as those based on stimulated Raman adiabatic passage (STIRAP) without a herald \cite{Specht2011}, could not be achieved.\par

By using a novel experimental apparatus, we demonstrate a scheme which overcomes these limitations, realising a passive, heralded quantum memory with high fidelity. Our scheme can be motivated from different perspectives. One view is based on the concept of impedance matching \cite{Trautmann2016}. Depending on the exact protocol, a photon impinging onto a cavity can only be stored efficiently if it is not reflected from the input mirror. This is the case when all other losses inside the cavity equal the transmission of the input mirror. For single-sided cavities, this requires additional losses which have to be induced by the atom in the course of the storage process. In case of STIRAP-based storage schemes, this is accomplished with suitably timed control fields. In our case, it is achieved in a completely passive manner via vacuum-stimulated Raman scattering into the mode of a second cavity. Detection of a photon at the output of that cavity heralds successful storage. Another view on our memory scheme follows the concept of vacuum-induced transparency \cite{Tanji2009}. Here, the vacuum field of an optical cavity aligned perpendicular to the propagation direction of an incoming light pulse makes the otherwise absorbing atomic medium transparent. This is accompanied by light scattering into the cavity. Absorption happens when the scattered light escapes from the cavity, thus heralding successful storage. By adding another cavity along the propagation direction of the incoming light, we bring this approach to the quantum regime, making the memory suitable for operation at the single atom and single photon level.\par

\setcounter{figure}{0}
\renewcommand{\figurename}{\textbf{Fig.}}
\renewcommand{\thefigure}{\textbf{\arabic{figure}}}
\makeatletter
\renewcommand{\@caption@fignum@sep}{ \textbar\ }
\makeatother

\begin{figure*}[!htp]
\centering
\makebox[\textwidth][c]{\includegraphics{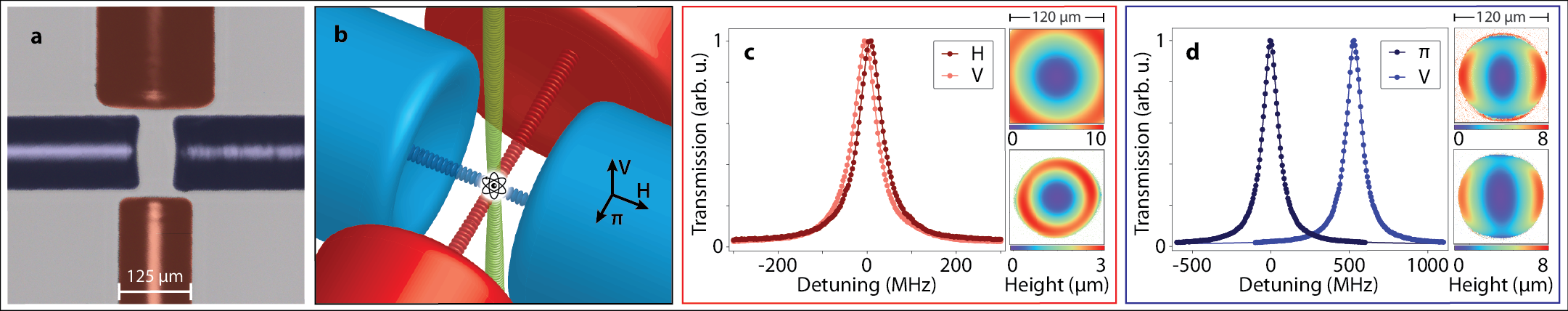}}
\caption{\label{fig:CrossedCavities}
\textbf{Crossed optical fibre cavities.} \textbf{a}, Coloured photograph of the crossed fibre cavities. Coloured in red is the qubit cavity, coloured in blue the herald cavity. \textbf{b}, 3D schematic of the crossed fibre cavities. The transparent beams illustrate the three-dimensional optical lattice which is used to trap single atoms (shown in black) at the crossing point of the cavity modes. \textbf{c, d}, Transmission spectra of the empty fibre cavities along with height maps of their constituting fibre tip surfaces, for the qubit cavity (\textbf{c}) and the herald cavity (\textbf{d}). Spectra in different colours represent different input polarisations, that are chosen along the eigenpolarisations of the cavities. The strongly elliptical mirror surfaces of the herald cavity lead to a large frequency splitting of the polarisation eigenmodes. Statistical error bars are smaller than the data points, solid lines are Lorentzian fits to the data.}
\end{figure*}

The heart of our apparatus are two single-sided fibre Fabry-Pérot cavities \cite{Hunger2010}, based on CO$_2$ laser-machined mirrors, crossing each other at an angle of 90° (Fig.\,\ref{fig:CrossedCavities}a). We refer to the cavity through which photonic qubits enter and exit the system as the qubit cavity, and it is shown in red in Fig.\,\ref{fig:CrossedCavities}a. The qubit cavity has spherical mirror surfaces to ensure polarisation-independent resonance frequencies (Fig.\,\ref{fig:CrossedCavities}c), and left-circular (L) and right-circular (R) polarisations are used as the polarisation basis for the qubit cavity, whose symmetry axis defines the quantisation axis throughout this paper. The cavity shown in blue in Fig.\,\ref{fig:CrossedCavities}a is used to create and collect herald photons during the storage process, and we refer to it in the following as the herald cavity. It was intentionally fabricated with strongly elliptical mirror surfaces, leading to two non-degenerate linear polarisation eigenmodes \cite{Uphoff2015} (Fig.\,\ref{fig:CrossedCavities}d), one of which is aligned to coincide with the $\pi$-polarisation of our system. This allows us to selectively couple the herald cavity to atomic transitions exclusively with the $\pi$-polarisation, which is orthogonal to the qubit cavity polarisations. Even though the frequency splitting of polarisation eigenmodes is large, the herald cavity is still perfectly suitable for operation with polarisation qubits in other protocols \cite{Duan2004}.\par

Single $^{87}$Rb atoms are trapped at the crossing point of the cavity modes inside a three-dimensional optical lattice (Fig.\,\ref{fig:CrossedCavities}b, see Methods for details). Depending on the atomic transition, both cavities show strong atom-photon coupling (see Supplementary Fig.\,\ref{fig:NormalModeSpectro}).\par

\begin{figure}[!htp]
\centering
\includegraphics{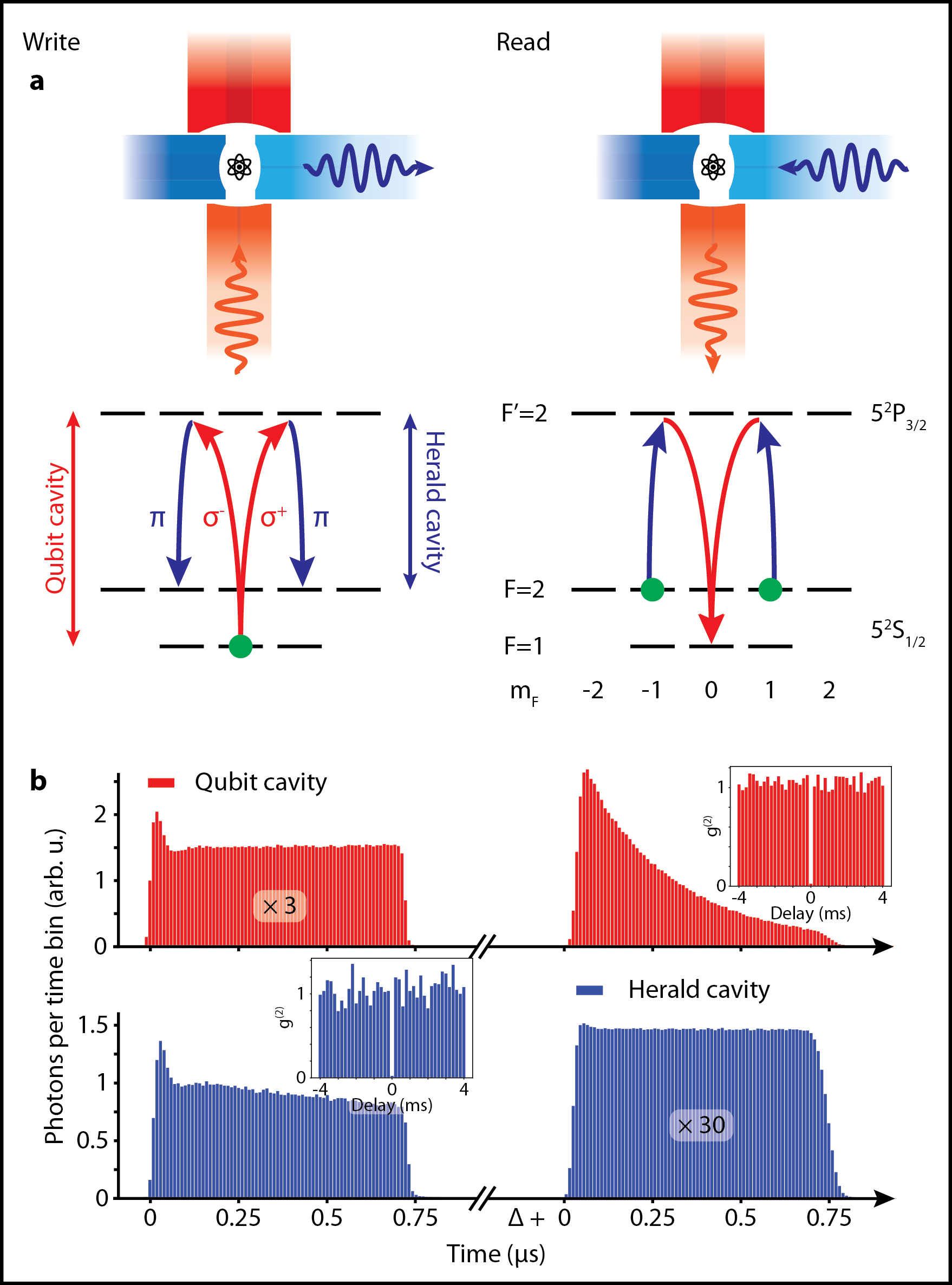}
\caption{\label{fig:StoreScheme}
\textbf{Scheme for the heralded quantum memory.} \textbf{a}, Schematic of the storage scheme. The left side illustrates the write process, the right side the read-out process. The upper part shows a drawing of the crossed cavities along with incoming and outgoing photons (wiggly arrows). The lower part shows the atomic level scheme. Green circles indicate initial atomic populations while curved arrows show the transitions that occur during the write and read-out process. \textbf{b}, Histograms of photons sent onto the cavities and leaving the cavities during the write process (left side) and read-out process (right side). The shape of a photon leaving a cavity reflects the shape of the coherent pulse sent onto the other cavity multiplied by the decaying probability for the atom to be still found in the initial state. The insets show g$^{(2)}$ correlation functions of herald photons (lower left) and the read-out photons (upper right). The strong suppression of correlations at equal times is a signature of single photons.}
\end{figure}

We operate the heralded quantum memory on the D$_2$ line of rubidium ($5^2S_{1/2}$ $\leftrightarrow$ $5^2P_{3/2}$) with the qubit cavity locked to the $F=1$ $\leftrightarrow$ $F'=2$ transition and the $\pi$-polarisation mode of the herald cavity locked to the $F=2$ $\leftrightarrow$ $F'=2$ transition (Fig.\,\ref{fig:StoreScheme}a). To characterise the memory, we use weak coherent laser pulses, containing $\bar{n}$ $\approx$ 0.5 photons on average as our photonic qubit input states where the polarisation is used to encode the input qubit $|\psi_{in}\rangle = \alpha|R\rangle + \beta|L\rangle$. The experiment starts by optically pumping the atom to the $|F=1, m_F=0\rangle$ initial state, and subsequently, sending a photonic qubit into the qubit cavity. This leads, ideally, to vacuum-stimulated emission of a photon into the herald cavity and the simultaneous transition of the atom to a final state where the qubit is encoded in a superposition of Zeeman states:
\begin{widetext}
\begin{align}
&\left(\alpha|R\rangle + \beta|L\rangle\right)_{\textrm{Input}}|F=1, m_F=0\rangle_{\textrm{Atom}} \rightarrow\nonumber\\
 &\hspace{1cm}\left(  \alpha|F=2, m_F = +1\rangle + \beta|F=2, m_F=-1\rangle \right)_{\textrm{Atom}}|1\rangle_{\textrm{Herald}}.\nonumber
\end{align}
\end{widetext}
Due to the strong birefringence, only the desired $\pi$-polarised decay is enhanced by the herald cavity. Detection of the emitted photon heralds successful storage, and after some variable storage time, the atomic state is read-out again. To that end, a $\pi$-polarised laser pulse ($\bar{n} \approx 6$) is sent onto the herald cavity, which inverts the write process and leads to the emission of a photon into the qubit cavity. The polarisation of this re-emitted photon is then analysed.\par

\begin{figure*}[!htp]
\centering
\makebox[\textwidth][c]{\includegraphics{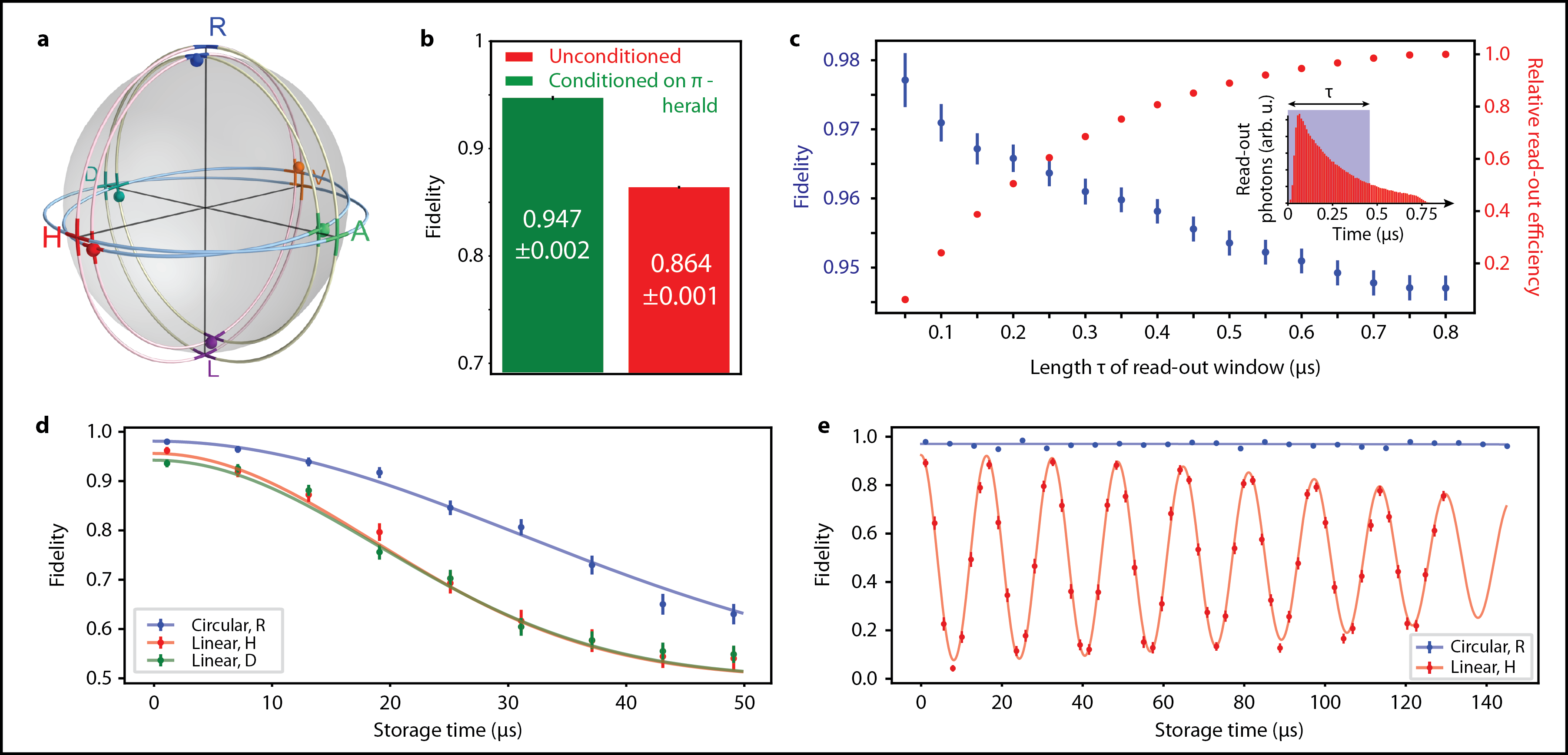}}
\caption{\label{fig:FidelityOfHQM}
\textbf{Fidelity of the heralded quantum memory.} \textbf{a}, Poincaré sphere showing the quantum process underlying the memory for an effective storage time of 1.1\,\textmu s. It is reconstructed using a maximum likelihood fit. The coloured spheres are the results of state tomography of the memory output for the correspondingly coloured and labelled input polarisations. The average state fidelity is (94.7$\pm$0.2)\,\%. \textbf{b}, Average state fidelity of the read-out photons with (green bar) and without (red bar) conditioning on the preceding detection of a herald photon, showing how the heralding improves the fidelity of the memory. \textbf{c}, Average state fidelity and relative read-out efficiency for truncated read-out photons. Disregarding late read-out photons increases the fidelity due to a reduced probability for the atom to have scattered a photon prior to the emission of a photon into the qubit cavity. \textbf{d}, Fidelity of the memory for longer storage times without magnetic guiding field. The fidelity decays over time, reaching the classical limit at storage times $\geq$~25\,\textmu s. \textbf{e}, Fidelity of the memory with an applied magnetic guiding field. While the fidelity of circular polarisation input states stays constantly high over time, the fidelity for linear input shows the expected oscillations, with an envelope reaching the classical limit at storage times $\geq$~170\,\textmu s. Error bars in \textbf{b}, \textbf{c}, \textbf{d}, \textbf{e} indicate the 1$\sigma$ confidence levels accounting for statistical uncertainties due to the finite number of detected photons.}
\end{figure*}

Figure~\ref{fig:FidelityOfHQM}a shows the result of a characterisation of our memory using state tomography of the read-out polarisation states for all six input polarisations along the coordinate axes of the Poincaré sphere with an effective storage time of 1.1\,\textmu s. Using post-selection on events where a herald photon was detected during the preceding write pulse, we find an average state fidelity $\bar{\mathcal{F}}_s$ of (94.7$\pm$0.2)\,\%, far beyond the classical limit of 69$\,\%$ for the applied coherent input. Alternatively, an underlying quantum process can be deduced from the data using a maximum likelihood fit, leading to a process fidelity $\mathcal{F}_p$ of (92.2$\pm$0.3)\,\%, in perfect agreement with the average state fidelity ($\bar{\mathcal{F}}_s = \left(2\mathcal{F}_p + 1\right)/3$).\par

When the storage time between write and read is extended, the fidelity drops, as expected, reaching the above mentioned classical threshold after $\geq$~25\,\textmu s (Fig.\,\ref{fig:FidelityOfHQM}d). We attribute this to residual, uncompensated magnetic fields, which lead to uncontrolled rotation of the atomic spin. By applying a magnetic guiding field of 44\,mG along the quantisation axis (Fig.\,\ref{fig:FidelityOfHQM}e), the coherence time of the memory is significantly improved. While the fidelity of circularly polarised input states, which are stored in energy eigenstates of the atom, is constantly high over the tested time interval, the fidelity of linearly polarised input states shows the expected oscillation at twice the Larmor frequency of 62\,kHz, with a decay time to the classical threshold of $\geq$~170\,\textmu s. Cooling the atoms to their motional ground state, mapping the atomic states to a decoherence-protected basis during storage \cite{Koerber2018}, or application of dynamical decoupling schemes \cite{Viola1999} are possible means to further improve the coherence time.\par

For short storage times, a number of causes are expected to contribute to the deviation of the memory from perfect fidelity, one of them being erroneous initialisation of the atomic state. And yet, part of the preparation errors are filtered out in the presented scheme as they will not lead to the emission of a herald photon. If one compares the fidelity of the memory with and without conditioning on the detection of a $\pi$-polarised herald photon, one finds that the conditioning improves the average state fidelity from (86.4$\pm$0.1)\,\% to (94.7$\pm$0.2)\,\% (Fig.\,\ref{fig:FidelityOfHQM}b). This reveals another advantage of the herald photon in the presented memory scheme, namely its capability to detect and filter out certain errors during the storage process, leading to an improved fidelity. A further improvement of the fidelity can be achieved by disregarding photons that are emitted at a late stage of the read-out process, leading to an average state fidelity reaching (97.7$\pm$0.4)\,\%, at the cost of a reduced read-out efficiency (Fig.\,\ref{fig:FidelityOfHQM}c). The reason behind this is a finite probability for the atom to scatter a photon and return to the $F=2$ hyperfine state prior to making the intended transition to the $F=1$ hyperfine state, which could be further mitigated by a higher cooperativity of the qubit cavity.\par

To quantify the efficiency of the heralded memory, we define the single-photon heralding efficiency as the probability to detect a herald photon for a single photon sent onto the qubit cavity. For the presented memory, this efficiency is (11$\pm$1)\,\%. It is reduced by a number of factors (see Methods), none of which is fundamental. The simplest way to improve the efficiency is to use state-of-the-art single photon detectors, which would nearly double the heralding efficiency. Further improvements can be expected from an increased atom-photon coupling, which can be achieved by improving the localisation of the atoms or by choosing different atomic states for performing the memory, such as having both the qubit and the herald cavity on the $F=1 \leftrightarrow F'=1$ transition of the D$_2$ line of $^{87}$Rb.

The atomic states chosen here have the advantage that a successful read-out process detunes the atom from the read-out light, which allows for efficient read-out of the qubit from the memory. We denote the read-out efficiency as the probability for retrieving a photon in the fibre in front of the qubit cavity provided that the atom was transferred to the $F=2$ state during the preceding write process. For the presented measurements, the read-out efficiency was (56$\pm$6)\,\% (see Methods for details).\par

For the measurements shown in Fig.\,\ref{fig:StoreScheme} and Fig.\,\ref{fig:FidelityOfHQM}, quasi-rectangular laser pulses were used during the write and read-out process. The same measurements were carried out with more smoothly shaped laser pulses, without any significant difference in the performance of the memory (see Supplementary Fig.\,\ref{fig:PulseShapes} and Fig.\,\ref{fig:FidelitiesShaped}). This is to be expected as the bandwidth of the quasi-rectangular pulses, which is limited by the switching time of the used acousto-optic modulators, is significantly below the bandwidth of the memory, which is on the order of the linewidth of the qubit cavity ($\approx 60$\,MHz). This illustrates the robustness of the presented storage scheme which, apart from centre frequency and bandwidth, does not put any requirements onto the impinging photon.\par

\begin{figure}[!htp]
\centering
\includegraphics{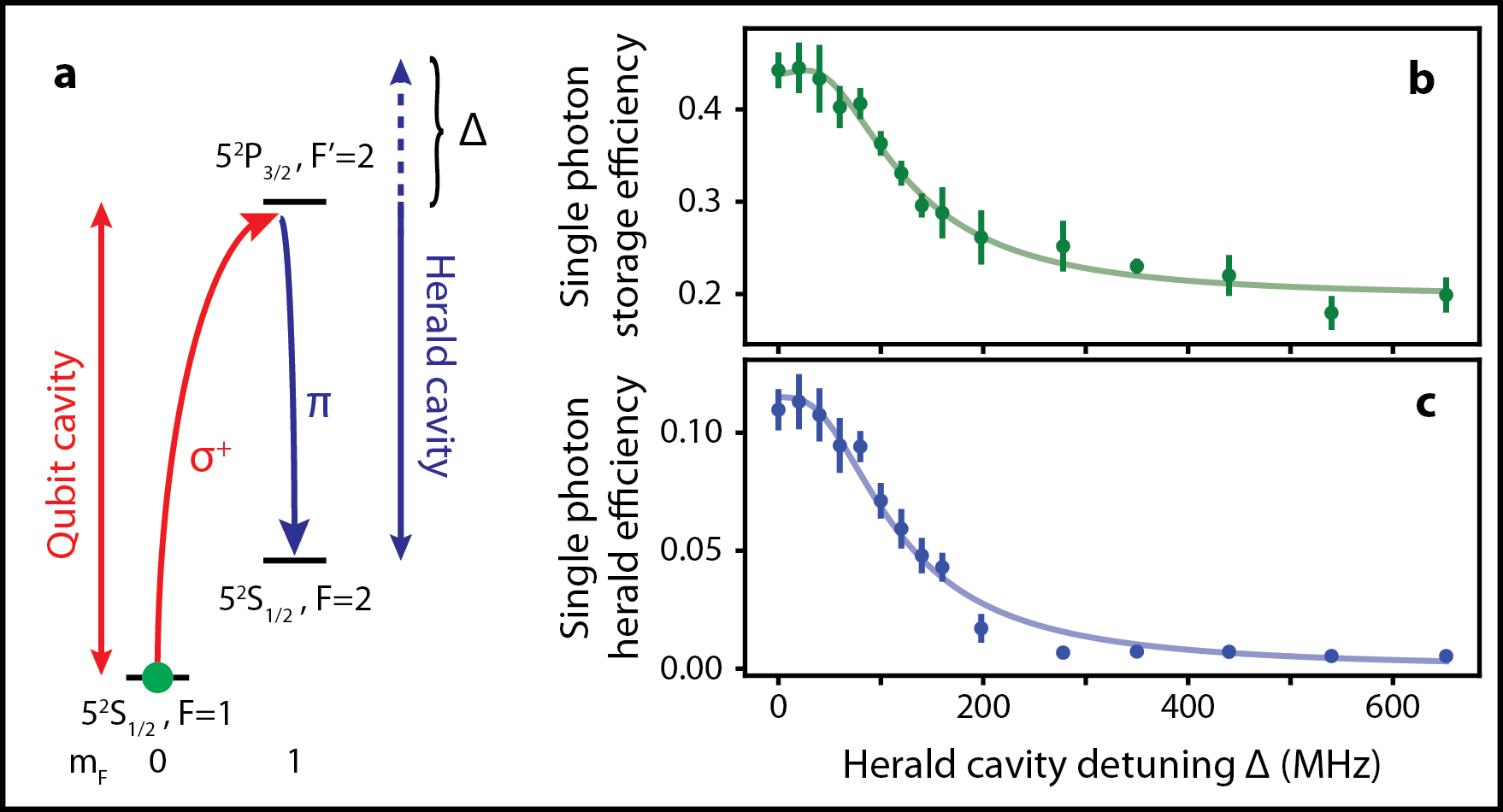}
\caption{\label{fig:EffVsKCDetuning}
\textbf{Heralded storage with variable herald cavity detuning.} \textbf{a}, The heralded storage experiment was performed with variable detuning of the herald cavity, as indicated by the dashed arrow. The green circle indicates the initial atomic population while the curved arrows show the transitions that occur during the heralded storage process. \textbf{b}, The probability that a single photon sent onto the qubit cavity transfers the atom to the final $5^2S_{1/2}, F=2$ state is shown for different herald cavity detunings. The decay of the storage efficiency with increasing herald cavity detuning shows that the storage process is induced by the vacuum field of the herald cavity. The error bars indicate the standard deviation over all atoms contributing to the measurements. The solid line shows a simple model describing the storage process, fitted to the data (see Methods for details). \textbf{c}, Analogous to \textbf{b}, this time showing the heralding efficiency for a single photon sent onto the qubit cavity. The detection efficiencies are included.}
\end{figure}

In order to determine the role of the vacuum-field of the herald cavity in the emission of a herald photon, we carried out another series of measurements where the frequency of the herald cavity was detuned relative to the atomic $F=2$ $\leftrightarrow$ $F'=2$ transition (Fig.\,\ref{fig:EffVsKCDetuning}a). The measurement results for both the single-photon storage efficiency and the single-photon heralding probability as a function of the herald cavity detuning are shown in Fig.\,\ref{fig:EffVsKCDetuning}b,c. The decrease of the transfer probability for increasing herald cavity detuning shows that qubit storage is predominantly induced by the vacuum-field of the herald cavity. The process can be well described using a simple model, based on the generic situation of cavity QED with a single atom coupled to the qubit cavity where the atom is subject to additional, Purcell-enhanced loss due to its coupling to the herald cavity (see Methods for details). Fit parameters of the model fitted to the single photon storage efficiencies shown in Fig.\,\ref{fig:EffVsKCDetuning}b,c are heuristic factors, by which the atom-photon coupling rates of the two cavities deviate from their expected values (see Methods). We attribute these to imperfections in the localisation of the atoms within the cavity modes and expect to be able to improve this in the future, for example by direct imaging of the atoms with a camera and subsequent feed-back onto the atomic position. This holds potential for further increase of the efficiency of the presented heralded quantum memory.\par

To conclude, we have presented a passive quantum memory network node where a herald photon is generated during qubit storage via a vacuum-induced Raman transition. Its robustness with respect to the shape and arrival time of the incoming photon are very useful features for applications in long distance fibre networks and in particular for the implementation of a quantum repeater \cite{Briegel1998, Uphoff2016}. The combination of robustness and high bandwidth due to the use of fibre cavities might also enable the storage of photons emitted by quantum dots \cite{Meyer2015}, thereby promoting the development of hybrid quantum systems. Last but not least, the crossed fibre cavities constitute a new platform for experiments on basic cavity QED \cite{Yoo1985}.

\vspace{0.5cm}
\newpage

\section*{Methods}
\subsection*{Crossed optical fibre cavities}
At the heart of the experimental apparatus are two crossed optical fibre cavities, whose constituting mirrors were shaped on the end facets of optical fibres using CO$_2$ laser-machining \cite{Hunger2010, Uphoff2015}. Each cavity consists of one highly-reflective mirror and one outcoupling mirror. The outcoupling mirrors were shaped on the end facets of single-mode fibres (\textit{IVG Fiber, Cu800}), for the highly-reflective mirrors multi-mode fibres were used (\textit{IVG Fiber, Cu50/200} and \textit{Cu50/125}) for the qubit cavity and the herald cavity, respectively. CO$_{2}$ laser-machining was done using a CO$_{2}$ laser emitting at a wavelength of 9.3\,\textmu m. The laser beam was focused to the centre of the fibres with a beam waist of 120--400\,\textmu m. For the machining of the elliptical mirrors, a beam with elliptical profile was used (ratio of the beam waists w$_y$/w$_x$ $\approx$ 2.2). The mirrors were shaped using 1--4 pulses with pulse lengths of 0.3--3.3\,ms and a laser power of 50\,W. 
The resulting radii of curvature of the fibre mirrors are roughly 340\,\textmu m and 170\,\textmu m for the outcoupling  and highly-reflective mirror of the qubit cavity and (R$_x$,\,R$_y$) = (100\,\textmu m,\,290\,\textmu m) and (90\,\textmu m,\,230\,\textmu m) for the outcoupling and highly-reflective mirror of the herald cavity, respectively. The exact values depend on the chosen region of interest as the mirror surfaces are not exactly spherical. The birefringent phase shift per round trip of the herald cavity that is expected due to the ellipticity of the mirrors \cite{Uphoff2015} (1.7\,mrad) is in good agreement with the direct measurement (1.8\,mrad).
The length of the qubit cavity and herald cavity are 162\,µm and 80\,µm, leading to expected mode waists at the cavity centre of w = 6.5\,µm and (w$_x$,\,w$_y$) = (3.5\,µm,\,4.8\,µm) for the qubit and herald cavity, respectively.
The shaped fibre end facets are coated with dielectric mirror coatings (\textit{Laseroptik GmbH}). The coating transmissions are 340\,ppm for the outcoupling mirrors and 10\,ppm for the highly-reflective mirrors. Additional losses due to absorption, scattering and mode-coupling lead to finesses of $14\,600\pm120$ and $15\,680\pm30$ for the qubit and herald cavity, respectively. No decay of the finesse was observed over time. In combination with the cavity lengths, this corresponds to field decay rates of $2\pi\cdot31.7$\,MHz and $2\pi\cdot59.8$\,MHz for the qubit and herald cavity, respectively.
The cavity fibres are mounted on stacks of piezo-electric slip-stick positioners that enable the adjustment of the cavity lengths, the mode-matching between cavity mode and fibre mode at the outcoupling mirrors, and the overlap of the modes of the two cavities in vacuum. The overlap of the modes is aligned using cross-correlation of fluorescence photons scattered into the cavities by atoms falling through the cavities to a precision of about 1\,\textmu m.

\subsection*{Single atoms in crossed cavities}
In order to bring a single atom into the cavities, we start by loading a cloud of $^{87}$Rb atoms into a magneto-optical trap (MOT) that is located about 10\,mm above the cavities. After a MOT loading phase (3\,s), a short phase of molasses cooling is applied that cools the atoms to a temperature of about 16\,\textmu K before the atoms are released and begin to fall. Shortly afterwards, a red-detuned standing-wave optical dipole trap ($\lambda=797.1$\,nm) is switched on that guides the falling atoms towards the crossing point of the cavity modes. An additional cooling beam (lin $\perp$ lin) that passes through the crossing point of the cavity modes under an oblique angle, induces cooling forces that stop some of the atoms when they arrive in the cavities. The atoms are trapped in a three-dimensional optical lattice ($U_0/k_B\approx$ 1\,mK), consisting of the above mentioned red-detuned standing-wave trap and two blue-detuned intracavity dipole traps ($\lambda=776.5$\,nm and 774.6\,nm). Light scattered into the cavities during cooling of the atoms is used to detect their presence inside the cavities. We use minimum bounds on the detected cooling light level to select atoms that are well placed within the cavity modes. Evaluation of g$^{(2)}$ correlations functions is used to ensure that only events where single atoms were trapped are considered for data analysis. The mean lifetime of atoms in the trap is typically on the order of 30\,s.

\subsection*{Heralded storage with subsequent read-out}
The measurements for the experiments on heralded storage with subsequent read-out were run with a repetition rate of 5\,kHz (for the tomography with short storage times) and 3.3\,kHz (for the characterisation of the coherence times). Atom cooling for about 140\,\textmu s was followed by 45\,\textmu s of optical pumping to initialise the atom in the $|F=1, m_F=0\rangle$ state. In order to further reduce residual population in the $|F=1, m_F= \pm1\rangle$ states, the last 4\,µs of optical pumping were done using $\pi$-polarised light resonant to the $F=1\rightarrow F’=1$ transition only. This is expected to come at the cost of some additional population in the $F=2$ states, which, however, does not emit any herald photons.  Subsequently, a write pulse with variable polarisation (duration $\tau\approx$ 740\,ns, average photon number $\bar{n}\approx$ 0.5) was sent onto the qubit cavity. After a certain storage time, the atomic state was read-out by sending a $\pi$-polarised read pulse ($\tau\approx$ 730\,ns, $\bar{n}\approx$ 6) onto the herald cavity. The effective storage times were finely calibrated using the starting point of the oscillation of the memory fidelity for linearly polarised input states when a magnetic guiding field was applied along the axis of the qubit cavity. An additional short sequence of pumping the atom to $F=1$ followed by a strong pulse ($\bar{n}\approx$ 10) sent onto the qubit cavity had the main purpose of getting g$^{(2)}$ correlation functions with higher signal-to-noise ratio in order to better discriminate between single and multiple trapped atoms, but was also used as a reference for efficiency calibration. \par

Polarisation tomography setups that allowed for the polarisation-resolved detection of light leaving the cavities or being reflected from their outcoupling mirrors were placed in front of both the qubit and the herald cavity. While the setup in front of the herald cavity was mainly used to ensure that the herald photons were $\pi$-polarised, the setup in front of the qubit cavity was used for full polarisation tomography of the photons read-out of the memory. 
Error bars and confidence intervals of the state fidelity $\mathcal{F}_s=\langle\psi_{in}|\rho_{out}|\psi_{in}\rangle$ of the memory output $\rho_{out}$ with respect to the input state $|\psi_{in}\rangle$ account only for the statistical uncertainties of $p_\parallel=N_\parallel/(N_\parallel+N_\perp)$ and follow the analysis of a Bernoulli process using normal approximation and 68\% confidence levels. Here, $N_\parallel$ ($N_\perp$) is the number of counts detected with polarisation parallel (orthogonal) to the input polarisation.
The stated uncertainty of the average state fidelity $\bar{\mathcal{F}_s}$ follows Gaussian uncertainty propagation.  
The process fidelity $\mathcal{F}_p$ is derived from a maximum-likelihood fit of a quantum process \cite{Specht2011} to the set of photon counts \{$N_{ij}|\, i,j\in\{R,L,H,V,A,D\}$\} acquired during the tomography measurements, where $i$ refers to the input polarisation and $j$ to the detection basis. Its uncertainty was assessed with a Monte Carlo method, using sets of numbers \{$\tilde{N}_{ij}$\}, where the $\tilde{N}_{ij}$  were randomly sampled from normal distributions centred at $N_{ij}$ with standard deviation $\sqrt{N_{ij}}$. The standard deviation of the resulting \{$\tilde{\mathcal{F}}_p$\} is used as the confidence interval for $\mathcal{F}_p$. \par

The probability $p_{H,1}$ to detect a herald photon for a single photon sent onto the qubit cavity was calculated from the probability $p_{H,\bar{n}}$ to detect a herald photon when a coherent laser pulse with $\bar{n}$ photons on average is sent onto the qubit cavity as
\begin{equation}
p_{H1} = -\frac{p_{H,\bar{n}}}{\bar{n}}\frac{\ln(1 - p_{t, \bar{n}})}{p_{t, \bar{n}}},\label{eq:ph1Fromphnptn}
\end{equation}
where $p_{t, \bar{n}}$ is the probability to transfer the atom from the $F=1$ to the $F=2$ hyperfine ground state with the coherent pulse sent onto the qubit cavity. Equation \eqref{eq:ph1Fromphnptn} uses the Poissonian photon number statistics of the coherent input pulse and the fact that there is no dark state other than $F=2$. The probability $p_{t, \bar{n}}$ was calculated from the ratio of herald counts detected during the write pulse divided by the number of counts detected during a much stronger reference pulse that to good approximation transferred all of the population. It can alternatively be assessed from the probability to detect a photon during the read-out process compared to the probability to detect a read-out photon conditioned on a preceding herald event, which was used to check for consistency. \par

The heralding efficiency for the measurements with rectangular input pulses of (11$\pm$1)\,\% is composed of (52$\pm$3)\,\% single-photon transfer efficiency from the initial $F=1$ state to the final $F=2$ state, (79$\pm$5)\,\% probability for emission of a $\pi$-polarised photon into the herald cavity during that process, (85$\pm$1)\,\% probability for the photon to leave the cavity through the mirror leading to the detector, (80$\pm$5)\,\% mode matching between the elliptical cavity mode and guided mode of the cavity fibre, (75$\pm$5)\,\% transmission efficiency to the detector and (50$\pm$4)\,\% detector efficiency.\par

The read-out efficiency of (56$\pm$6)\,\% is calculated from the probability to detect a read-out photon subsequent to a heralded storage event. It consists of (92$\pm$1)\,\% probability for the atom to be transferred to the $F=1$ state, (86$\pm$6)\,\% probability to emit a photon into the qubit cavity during that process, (79$\pm$1)\,\% probability for the photon to leave the cavity through the right mirror and (90$\pm$5)\,\% mode matching between cavity mode and fibre mode.\par

The stated uncertainties for the efficiencies and related parameters are estimated from the spread of parameters obtained for different measurements during the tomography and further account for systematic uncertainties of certain experimental parameters, like the input photon number and detector efficiencies, if relevant. The spread of evaluated parameters for different measurements is believed to mainly reflect the spread of trapping sites of atoms within the cavity modes, but can additionally be due to drifts of parameters, such as mode matchings, during the experiment.

\subsection*{Heralded storage for variable herald cavity detuning}
The experimental sequence for the measurements on heralded storage with variable detuning of the herald cavity was run at a repetition rate of 1\,kHz and consisted of three subparts, each lasting for 300\,\textmu s. The first subpart started by 230\,\textmu s of cooling and optically pumping the atom to the $|5^2S_{1/2}, F=1, m_F=0\rangle$ state, before a weak coherent write pulse ($\bar{n}\approx$ 0.5, circular polarisation, duration about 800\,ns) was sent onto the qubit cavity. Instead of subsequently mapping the atomic state back to a photon, cavity-assisted fluorescence state detection of the $|5^2S_{1/2}, F=2\rangle$ state was carried out for about 60\,\textmu s. A short sequence of optically pumping the atom to the $|5^2S_{1/2}, F=1\rangle$ hyperfine state was added at the end to discriminate single from multiple trapped atoms using g$^{(2)}$ correlation functions. The second and third subpart of the sequence were variants of the first one and were used for reference. While the write pulse was replaced by a much stronger pulse ($\bar{n}\approx50$) that transferred all the population to the $|5^2S_{1/2}, F=2\rangle$ state in the second subpart, no pulse was sent onto the qubit cavity during the third subpart.\par

The probability $p_{t, \bar{n}}$ with which the atom is transferred from the $F=1$ to the $F=2$ hyperfine ground state when a coherent write pulse with $\bar{n}$ photons on average is sent onto the qubit cavity was derived by comparing the state detection signals acquired during the three subparts of the experimental sequence, averaged over all sequences run on a given trapped atom. Alternatively, the ratio of herald photons detected during the regular write pulse and the strong reference pulse as well as single-shot state detection, which was possible only for some herald cavity detunings, were used to check for consistency. Using the same assumptions as in Eq.\,\eqref{eq:ph1Fromphnptn}, the single photon storage efficiency $p_{s}$, i.e., the probability for the atom to be transferred from the $F=1$ to the $F=2$ hyperfine state when a single photon is sent onto the qubit cavity, is derived from $p_{t, \bar{n}}$ as
\begin{align*}
p_s =  -\frac{\ln(1-p_{t,\bar{n}})}{\bar{n}}.
\end{align*}
Single photon heralding efficiencies were calculated using Eq.\,\eqref{eq:ph1Fromphnptn}.\par

The data shown in Fig.\,\ref{fig:EffVsKCDetuning} represents the average and standard deviation of the set of all contributing atoms for a given detuning $\Delta$. The impact of different atoms was weighted by their storage time. Specifically, efficiencies were first calculated for all contributing atoms separately, yielding average values $\mu$ with statistical variance $\sigma^2$. These values were used to add as many normally distributed random numbers $X\sim\mathcal{N}(\mu,\sigma^2)$ to an overall set $S_{\Delta}$ as there were experimental sequences during the storage time of the atom. A Gaussian fit to $S_{\Delta}$ provided the efficiencies and standard deviations as shown in Fig.\,\ref{fig:EffVsKCDetuning}.

\subsection*{Model for the vacuum-induced heralded storage}

The model we use to describe the vacuum-induced storage process as shown in Fig.\,\ref{fig:EffVsKCDetuning} is based on an evaluation of possible loss channels for a photon impinging onto the qubit cavity. One fraction of the photon, $\mathcal{P}_{R}$, will be back-reflected from the qubit cavity, another fraction, $\mathcal{P}_{C}$, will be lost from within the cavity due to scattering and absorption from the cavity mirrors and transmission through the mirror at the back of the cavity, and a third fraction, $\mathcal{P}_{A}$, will be scattered by the atom. This last fraction will go along with a transfer of the atom the $F=2$ hyperfine ground state with probability $p_{F2}$, leading to a storage efficiency $p_s$ of: 
\begin{widetext}
\begin{align}
p_{s}(\gamma_{P}) = \mathcal{P}_{A}(\gamma_{P})\cdot p_{F2}(\gamma_{P}) = (1 - \mathcal{P}_{R}(\gamma_{P}) - \mathcal{P}_{C}(\gamma_{P}))\cdot \frac{\gamma/2 + \gamma_{P}}{\gamma + \gamma_{P}}
\label{eq:modelStorageProbab}
\end{align}
\end{widetext}
where $ \gamma_{P}$ is the vacuum-induced decay rate into the herald cavity,
\begin{align*}
\gamma_{P}(\Delta) = \frac{g_H^2\kappa_H}{\kappa_H^2+(2\pi\Delta)^2},
\end{align*}
which is a function of the detuning $\Delta$ between the atom and the herald cavity (see Fig.\,\ref{fig:EffVsKCDetuning}a). Here, $g_H$ and $\kappa_H$ are the atom-cavity coupling rate and the field decay rate of the herald cavity, respectively. 

The fractions $\mathcal{P}_{R}$ and $\mathcal{P}_{C}$ are deduced from the reflection and transmission coefficients of the qubit cavity, as calculated using the input-output formalism under the assumption of a weak cavity drive. The effect of the herald cavity enters via the Purcell-enhanced decay of the atom $\gamma\mapsto\tilde{\gamma}=\gamma+\gamma_p$. Denoting $\epsilon = (2\tilde{\gamma}\sqrt{\kappa_Q\kappa_{1Q}})/(g_Q^2+\kappa_Q\tilde{\gamma})$, where $g_Q$ and $\kappa_Q$ are the atom-cavity coupling rate and the field decay rate of the qubit cavity, respectively, and $\kappa_{1Q}$ is the decay rate through the outcoupling mirror, we get:
\begin{widetext}
\begin{align*}
\mathcal{P}_{C} = \frac{\kappa_Q-\kappa_{1Q}}{\kappa_Q}\mu_{FC}^2\epsilon^2,\hspace{.5cm} \mathcal{P}_{R} = (1-\mu_{RC}^2) + \left|\mu_{RC} - \mu_{FC}\sqrt{\frac{\kappa_{1Q}}{\kappa_Q}}\epsilon\right|^2.
\end{align*}
\end{widetext}
Here, $\mu_{FC}$ is the amplitude mode overlap between the mode of the qubit cavity and the guided mode of the fibre in front of it, and $\mu_{RC}$ is the amplitude mode overlap between the mode of the qubit cavity and the mode to which the incoming fibre mode is mapped when reflected off the curved outcoupling mirror of the qubit cavity.\par

The fit parameters of the model to the single photon storage efficiencies shown in Fig.\,\ref{fig:EffVsKCDetuning}b are the mode matchings $\mu_{FC}^2$ and $\mu_{RC}^2$ as well as heuristic factors, by which the atom-cavity coupling rates are reduced compared to our expectations based on the cavity geometries. Their fitted values are $0.8$ for $\mu_{FC}^2$, $0.95$ for $\mu_{RC}^2$, and $0.6$ for the reduction factors of the atom-cavity coupling rates. The single photon heralding probability as shown in Fig.\,\ref{fig:EffVsKCDetuning}c follows from Eq.\,\eqref{eq:modelStorageProbab} by assuming that only photons emitted during Purcell-enhanced decay are collected by the herald cavity. The only remaining free fit parameter in Fig.\,\ref{fig:EffVsKCDetuning}c is an overall detection efficiency for a photon present in the herald cavity ($\eta$ = 0.3), all other parameters are used from the fit of the single photon storage efficiency (Fig.\,\ref{fig:EffVsKCDetuning}b).

\vspace{0.5cm}
\begin{sloppypar}
\noindent\textbf{Data Availability} The data that support the findings of this study are available from the corresponding author upon reasonable request.
\end{sloppypar}

\vspace{0.5cm}
\noindent\textbf{Acknowledgements} We thank Stephan Ritter and Manuel Uphoff for contributions during an early stage of this work and Tobias Urban for contributions to the design and fabrication of the experimental chamber. This work was supported by the Bundesministerium für Bildung und Forschung via the Verbund Q.Link.X (Grant No. 16KIS0870), by the Deutsche Forschungsgemeinschaft (DFG, German Research Foundation) under Germany's Excellence Strategy - EXC-2111 - 390814868 and the European Union's Horizon 2020 research and innovation programme via the project Quantum Internet Alliance (QIA, GA No. 820445). J.\,D.\,C. acknowledges support from the Alexander von Humboldt Foundation.

\vspace{0.5cm}
\noindent\textbf{Author Contributions} All authors contributed to the experiment, the analysis of the results and the writing of the manuscript.

\vspace{0.5cm}
\begin{sloppypar}
\noindent\textbf{Author Information} The authors declare that they have no competing financial interests. Correspondence and requests for materials should be addressed to M.\,B. (manuel.brekenfeld@mpq.mpg.de).
\end{sloppypar}

\onecolumngrid
\vspace{170pt}
\section{Supplementary Information}
\vspace{1.2cm}
\twocolumngrid

\setcounter{figure}{0}
\renewcommand{\figurename}{\textbf{Supplementary Fig.}}
\renewcommand{\theHfigure}{S.\arabic{figure}}

\setcounter{page}{1}
\renewcommand{\thepage}{\roman{page}}

\section*{Normal-mode spectroscopy}
In order to characterise the coupling of single trapped atoms to the two cavities, we performed normal-mode spectroscopy for both cavities separately. The cavity to be characterised was locked to the $F=2$ $\leftrightarrow$ $F'=3$ transition of the D$_2$ line of rubidium \mbox{($5^2S_{1/2}$ $\leftrightarrow$ $5^2P_{3/2}$)}, while the other cavity was detuned, to provide an optical lattice without being resonant to any atomic transition. The experimental sequence was run at a repetition rate of 4\,kHz, starting with 140\,\textmu s of cooling, followed by 40\,\textmu s of optical pumping, after which the cavity transmission was probed at varying detunings for 50\,\textmu s. A final stage of optical pumping between the two hyperfine ground states was used to ensure the presence of single atoms via the evaluation of g$^{(2)}$ correlation functions.\par
The measurements for the qubit cavity and for the herald cavity differed in details regarding the initialisation of the atomic state and the polarisation used to characterise the atom-cavity coupling. For the qubit cavity, the atom was initially pumped to the $|F=2, m_F=2\rangle$ state and the atom-cavity coupling was probed with right-circular light on the $|F=2, m_F=2\rangle\leftrightarrow|F'=3, m_F=3\rangle$ cycling transition, where the coupling strength is expected to be largest. For the herald cavity, the optical pumping and probing scheme were different, due to the fact that the herald cavity does not support circular polarisations. The quantisation axis was now set to point along the normal to the plane containing the two fibre cavities (indicated as direction V in Fig.\,\textbf{1}b in the main text). For the measurements shown below, the atom was initialised in the $|F=2, m_F=2\rangle$ state and the atom-cavity coupling was probed with $\pi$-polarised light on the $|F=2, m_F=2\rangle\leftrightarrow|F'=3, m_F=2\rangle$ transition. From the strength of the probe pulse, we expect that some of the atomic population is transferred to neighbouring Zeeman states during the probe interval.\par
Normal mode spectra for both the qubit and the herald cavity are shown in Supplementary Fig.\,\ref{fig:NormalModeSpectro}. Both spectra show data collected from one, single trapped atom. The data is fitted 
with an analytic formula for the cavity transmission$^{1}$. Fit parameters are the atom-cavity coupling rate and the atom cavity detuning. The cavity decay rates are taken from independent characterisations. The atom-cavity coupling rates for the data shown in Supplementary Fig.\,\ref{fig:NormalModeSpectro} are \mbox{$g=2\pi\cdot(36.7\pm0.3)\,$MHz} for the qubit cavity and \mbox{$g=2\pi\cdot(41.5\pm0.4)\,$MHz} for the herald cavity. Both values are about a factor of two below our expectations based on the geometric parameters of the cavities and fluctuate for different atoms, depending on the atom's exact position within the cavity modes. We expect to be able to improve the localisation of the atoms in the future, for example by focusing the optical dipole trap on the axis normal to the plane of the cavities more tightly.

\begin{figure*}[!htp]
\centering
\includegraphics{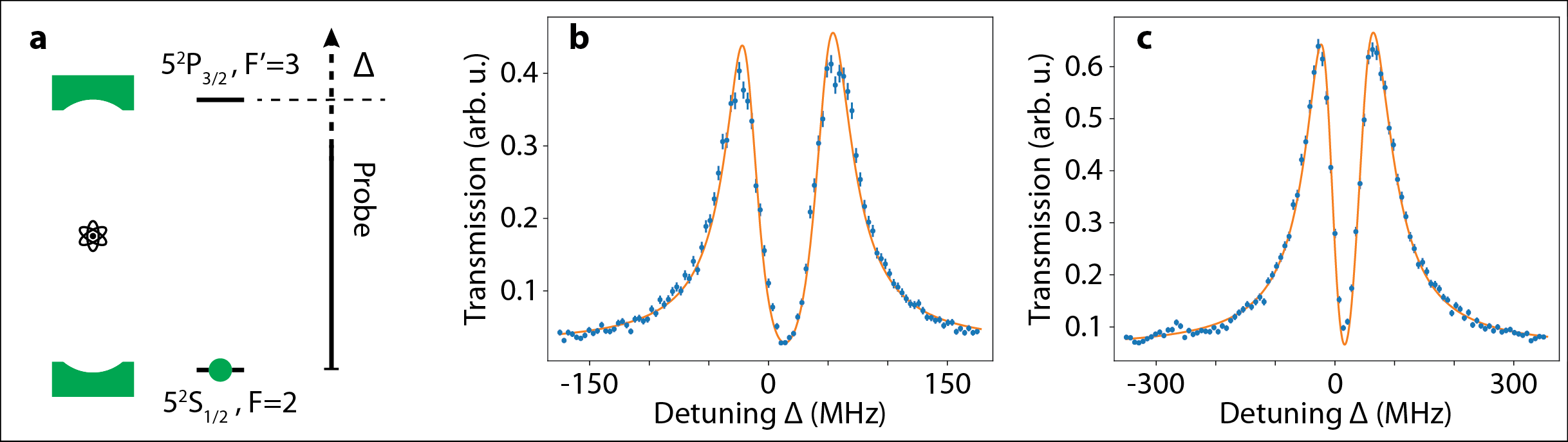}
\protect\caption{\textbf{Normal-mode spectroscopy.} Normal-mode spectroscopy of single atoms coupling to the qubit cavity (\textbf{b}) and to the herald cavity (\textbf{c}). Plotted is the transmission through either cavity as a function of the detuning relative to the  $|F=2\rangle\leftrightarrow|F'=3\rangle$ transition of an untrapped atom (\textbf{a}) . The error bars indicate statistical uncertainties assuming shot noise limitation. The solid, orange lines show the fit of an analytic model. See Supplementary Information for details on the experiment.}
\protect\label{fig:NormalModeSpectro}
\end{figure*}

\section*{Heralded storage with different pulse shapes}
In addition to the pulses with quasi-rectangular temporal profile shown in the main part of the paper, we also used more smoothly shaped pulses to characterise the heralded quantum memory. These pulses are shown in Supplementary Fig.\,\ref{fig:PulseShapes}. The average input photon number of the write pulse was again $\bar{n}\approx 0.5$, the average photon number of the pulse sent onto the herald cavity for read-out was $\bar{n}\approx6$. \par
The results for the fidelity, based on polarisation tomography of the read-out photons, are shown in Supplementary Fig.\,\ref{fig:FidelitiesShaped}a. The average state fidelity is $(94.1\pm0.1)\,\%$, the process fidelity is $(91.1\pm0.1)\,\%$, both very similar to the fidelities obtained for quasi-rectangular pulse shapes. Without conditioning on a previously detected herald photon, the average state fidelity is $(84.6\pm0.1)\,\%$ (Supplementary Fig.\,\ref{fig:FidelitiesShaped}b). The slightly lower fidelity values of the smooth pulses compared to the quasi-rectangular pulses might be due to a slightly longer effective storage time, resulting from the shifted centre of mass of the read-out photons, or it might be due to slightly different experimental parameters (e.g. compensation of external magnetic fields and polarisation drifts in optical fibres). The complete process matrices of the write-read process of the heralded memory are shown in Supplementary Fig.\,\ref{fig:ProcessMatrices}, for both the characterisation with smoothly shaped pulses (Supplementary Fig.\,\ref{fig:ProcessMatrices}a) and quasi-rectangular pulses (Supplementary Fig.\,\ref{fig:ProcessMatrices}b). \par
The single photon heralding efficiency is $(11\pm1)\,\%$, the single photon storage efficiency is $(55\pm4)\,\%$, which, within the measurement uncertainties, is in agreement with the values measured for the quasi-rectangular pulses.

\begin{figure*}
\centering
\includegraphics{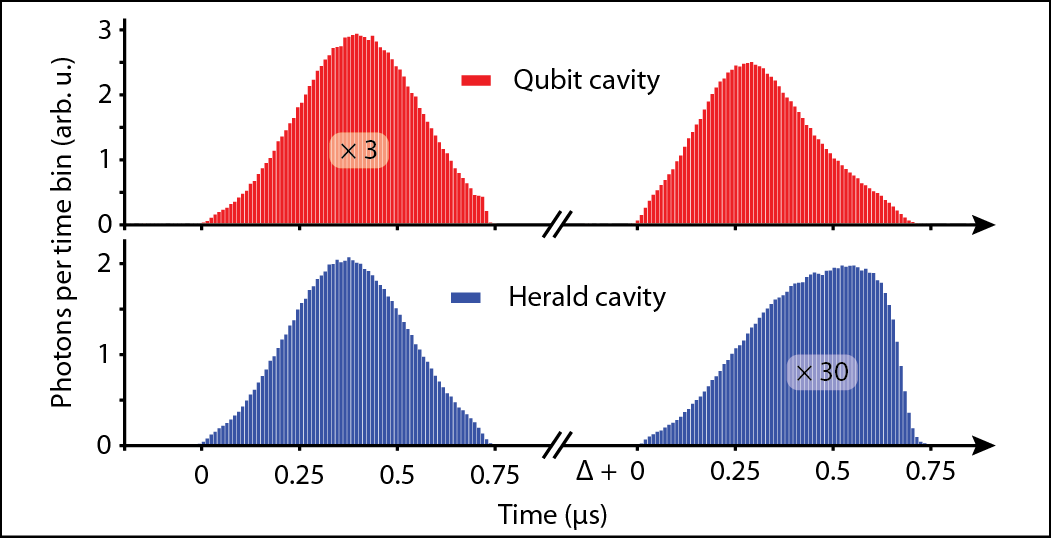}
\caption{\textbf{Pulse shapes used for complementary characterisation.} Pulse shapes for a complementary characterisation of the heralded quantum memory. Shown on the left side is the write process, at the top the photons sent onto the qubit cavity, at the bottom the photons exiting the herald cavity. Shown on the right side is the read-out process, at the bottom the read-pulse sent onto the herald cavity, at the top the read-out photons exiting the qubit cavity. See Supplementary Information for details.}
\label{fig:PulseShapes}
\end{figure*}

\begin{figure*}
\centering
\includegraphics{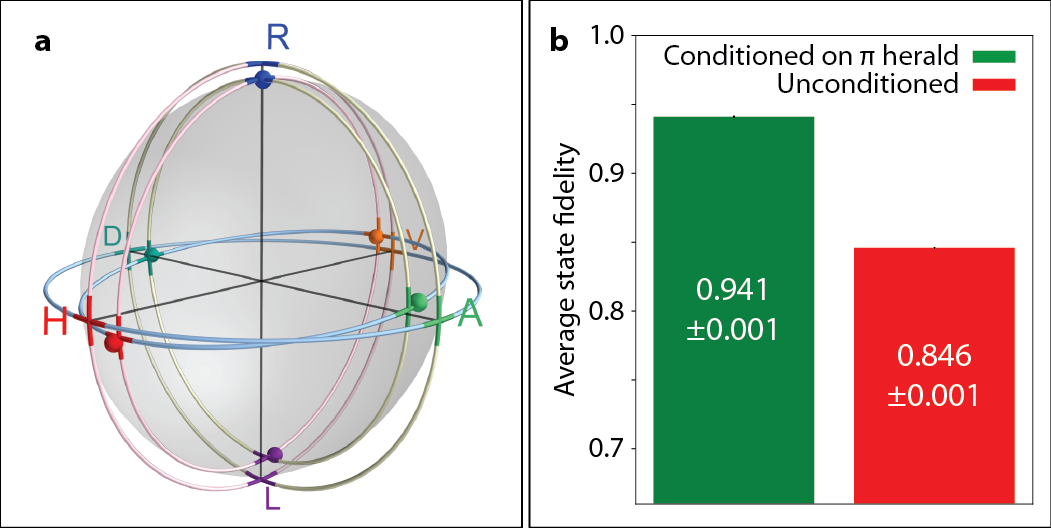}
\caption{\textbf{Fidelities of complementary characterisation.} \textbf{a}, Poincaré sphere showing the polarisations of the read-out photons. The sphere shows the underlying quantum process, reconstructed from  a quantum process tomography using a maximum likelihood fit. The coloured dots show, for each input polarisation, the results of a quantum state tomography of the memory output. \textbf{b}, Comparison of the fidelity of the quantum memory, depending on whether or not the read-out is conditioned on the preceding detection of a herald photon. See Supplementary Information and Methods for details.}
\label{fig:FidelitiesShaped}
\end{figure*}

\begin{figure*}
\centering
\makebox[\textwidth][c]{\includegraphics{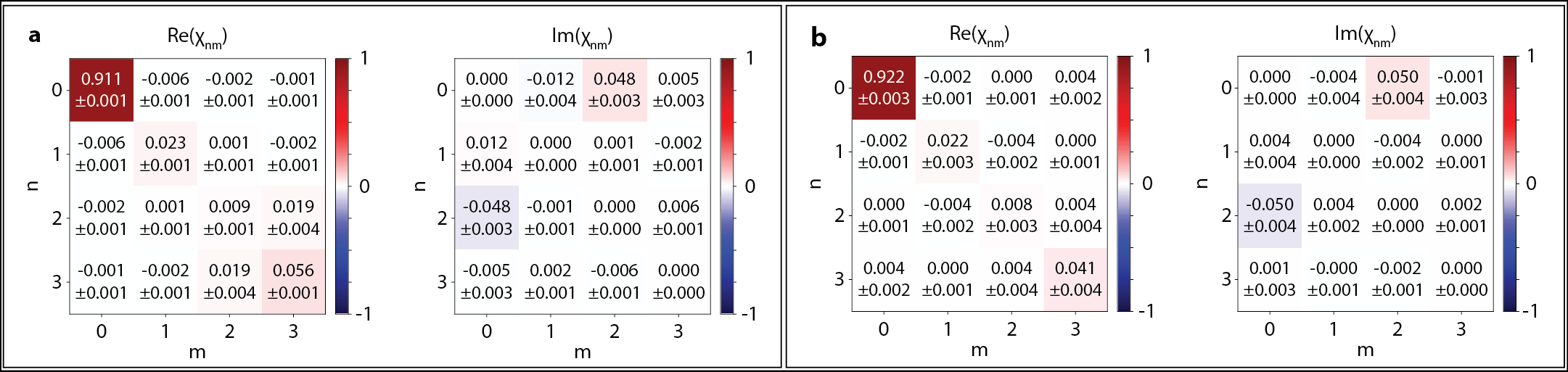}}
\caption{\textbf{Process matrices of the heralded quantum memory.} \textbf{a}, Process matrix of the write-read process of the heralded quantum memory for characterisation with smoothly shaped pulses. \textbf{b}, The equivalent process matrix for characterisation of the memory using the quasi-rectangular input pulses shown in the main text of this article. See Methods for details on the evaluation.}
\label{fig:ProcessMatrices}
\end{figure*}

\section*{References for Supplementary Information}
\noindent
$^{1.}$ Reiserer,~A. \& Rempe,~G. Cavity-based quantum networks with single atoms
and optical photons. {\it Rev. Mod. Phys.\/} {\bf 87}, 1379 (2015).

\end{document}